# PHASE SPACE FORMULATION OF QUANTUM MECHANICS. INSIGHT INTO THE MEASUREMENT PROBLEM

D. Dragoman* – Univ. Bucharest, Physics Dept., P.O. Box MG-11, 76900 Bucharest, Romania

**Abstract**:

A phase space mathematical formulation of quantum mechanical processes accompanied by and ontological interpretation is presented in an axiomatic form. The problem of quantum measurement, including that of quantum state filtering, is treated in detail. Unlike standard quantum theory both quantum and classical measuring device can be accommodated by the present approach to solve the quantum measurement problem.

* Correspondence address: Prof. D. Dragoman, P.O. Box 1-480, 70700 Bucharest, Romania, email: danieladragoman@yahoo.com

## 1. Introduction

At more than a century after the discovery of the quantum and despite the indubitable success of quantum theory in calculating the energy levels, transition probabilities and other parameters of quantum systems, the interpretation of quantum mechanics is still under debate. Unlike relativistic physics, which has been founded on a new physical principle, i.e. the constancy of light speed in any reference frame, quantum mechanics is rather a successful mathematical algorithm. Quantum mechanics is not founded on a fundamental principle whose validity may be questioned or may be subjected to experimental testing; in quantum mechanics what is questionable is the meaning of the concepts involved. The quantum theory offers a recipe of how to quantize the dynamics of a physical system starting from the classical Hamiltonian and establishes rules that determine the relation between elements of the mathematical formalism and measurable quantities. This set of instructions works remarkably well, but on the other hand, the significance of even its landmark parameter, the Planck's constant, is not clearly stated. Symptomatic for the poor understanding of the fundamentals of quantum mechanics is the existence of (at least) nine different formulations [1] (including the Heisenberg and Schrödinger formulations, the path integral formulation, the pilot wave and variational formulations), not to mention the different interpretations. (A list of essential bibliography for each formulation can be also found in [1].) Although classical mechanics can be as well formulated in many different ways (there is a Hamiltonian, a Lagrangian formulation, and so on), the different formulations of quantum mechanics are often based on different views about the physical reality.

Perhaps the most troublesome problem of quantum mechanics is measurement. There is a large number of theories that try to explain the apparent controversy between the superposition principle of quantum mechanics and the probabilistic results of measurement, i.e. to explain how to reconcile the occurrence with a certain probability of eigenvalues of a certain operator but exclude any evidence for the superposition of the operator's eigenstates in which the quantum state can mathematically be expanded. This discrepancy between mathematics and measurement results is still under scrutiny. The earliest attempt to solve it was made by Bohr (see the reprints in [2]) who draw a border between the quantum system, subject to the superposition principle, and the classical measuring device (including the observer), to which this principle does not apply. Another approach to the quantum measurement problem was offered by von Neumann [3] who included also the measuring apparatus in a quantum description but was forced to postulate the reduction or collapse of the quantum mechanical wavefunction in order to explain the measurement results. A more recent explanation of the wavefunction collapse is provided by the interaction of the quantum system with the environment, which induces the loss of phase coherence (decoherence) of the superposition between a set of preferred states singled out by the environment (see the review in [4]). The many-worlds interpretation of quantum mechanics [5], which assumes that at suitable interactions the wavefunction of the universe splits into several branches, or the superselection rules, which imply that the essence of a quantum mechanical measurement resides in the non-observation of a certain part of the system (see the review in [6]), are examples of other quantum mechanical theories that attempted to solve the measurement problem. The list of such attempts is much longer and still growing but as for many issues in quantum mechanics, including the problems of defining the time or phase, there is no agreement upon the theory of measurement.

The present paper offers another solution to this unsettled issue, which has the advantage of starting from a physical principle and a consistent interpretation of the measurement results, and not from another mathematical formalism. As will become clearer in the paper the phase space formulation of quantum theory is the most suitable for outlining the skeleton of this new interpretation of quantum mechanics and, in particular, of the measurement problem. The theory presented in this paper is laid down as a set of postulates that address not only the mathematical formulation but also the ontological viewpoint of the author about physical reality.

## 2. Postulates of the quantum theory

The quantum theory presented in this paper is based on a number of eight postulates.

**Postulate 1**: Quantum particles have physical reality.

Through quantum particles I understand localized and indivisible concentrations of energy (such as photons) and/or mass (electrons, etc.). The existence, evolution and interaction of quantum particles do not depend on the presence of an observer.

**Postulate 2**: The state of a quantum particle can only be described by a vector on the Hilbert space or by a quasi-probability distribution in the phase space formulation of quantum mechanics.

This postulate introduces nothing new in comparison to standard quantum mechanics; it only places the quasi-probability distribution on the same footing as the representation of the quantum state through a vector on the Hilbert space. To be more specific, the Wigner distribution function (WDF) is taken as the phase space quasi-probability distribution throughout this paper. For a quantum system with $n$ degrees of freedom characterized by a wavefunction $\Psi(\boldsymbol{q};t) = \Psi(q_1, q_2, ..., q_n; t)$ the WDF is defined as [7]

$$W(\boldsymbol{q}, \boldsymbol{p}; t) = \frac{1}{h^n} \int \Psi^* \left( \boldsymbol{q} - \frac{\boldsymbol{x}}{2}; t \right) \Psi \left( \boldsymbol{q} + \frac{\boldsymbol{x}}{2}; t \right) \exp(-i\boldsymbol{px}/\hbar) d\boldsymbol{x}$$

$$= \frac{1}{h^n} \int \overline{\Psi}^* \left( \boldsymbol{p} - \frac{\boldsymbol{r}}{2}; t \right) \overline{\Psi} \left( \boldsymbol{p} + \frac{\boldsymbol{r}}{2}; t \right) \exp(i\boldsymbol{qr}/\hbar) d\boldsymbol{r} \ . \tag{1}$$

where $\boldsymbol{qp}$ is a shorthand notation for $q_1 p_1 + q_2 p_2 + ... + q_n p_n$ and

$$\overline{\Psi}(\boldsymbol{p}) = h^{-n/2} \int \Psi(\boldsymbol{q}) \exp(-i\boldsymbol{pq}/\hbar) \tag{2}$$

is the Fourier transform of the quantum wavefunction. The phase space of quantum mechanics is spanned by the coordinate and momentum vectors on the classical phase space: $\boldsymbol{q}$ and $\boldsymbol{p}$, respectively. The WDF, as function of the classical phase space coordinates, carries the same information as the quantum state description through a vector on the Hilbert space. Moreover, it can also be defined for mixed states characterized by a Hermitian density operator $\hat{\rho}$ as

$$W(\boldsymbol{q}, \boldsymbol{p}; t) = \frac{1}{h^n} \int \left\langle \boldsymbol{q} + \frac{\boldsymbol{x}}{2} \middle| \hat{\rho}(t) \middle| \boldsymbol{q} - \frac{\boldsymbol{x}}{2} \right\rangle \exp(-i\boldsymbol{px}/\hbar) d\boldsymbol{x} , \tag{3}$$

being exactly analogous to the density matrix of Dirac and von Neumann [8]. Due to this remarkable property the WDF has long been used to develop an alternative formulation of quantum mechanics to that of Heisenberg and Schrödinger; for reviews on the properties and applications of the WDF as well as to its relations to other distribution functions see [9-12]. The real-valued quantum-mechanical WDF, although analogous to Gibbs phase space density [13], is not a true density in phase space since it can also take negative values. The occurrence of such "negative probabilities" has been interpreted as either the impossibility of simultaneously measuring conjugate variables such as position and momentum [8] or as an indication of phase space interference between different minimum phase space areas (quantum blobs in quantum mechanics [14]) covered by a given state [15].

The wavefunction can be recovered from the WDF as

$$\Psi(\boldsymbol{q})\Psi^*(\boldsymbol{0}) = \int W(\boldsymbol{q}/2, \boldsymbol{p}) \exp(i\boldsymbol{pq}/\hbar) d\boldsymbol{q} , \tag{4}$$

where $\boldsymbol{0} = (0,0,...,0)$, a similar relation existing for $\overline{\Psi}(p)$.

**Postulate 3**: The WDF is the quasi-probability of finding a quantum particle in phase space.

This postulate specifies the meaning of the WDF. From a mathematical point of view, since the squared modulus of the wavefunction and its Fourier transform are related to the WDF through

$$| \Psi(\boldsymbol{q}) |^2 = \int W(\boldsymbol{q}, \boldsymbol{p}) d\boldsymbol{p} , \qquad | \overline{\Psi}(\boldsymbol{p}) |^2 = \int W(\boldsymbol{q}, \boldsymbol{p}) d\boldsymbol{q} , \qquad (5)$$

it follows that

$$\int W(\boldsymbol{q}, \boldsymbol{p}) d\boldsymbol{q} d\boldsymbol{p} = 1 . \qquad (6)$$

However, since even the meaning of the quantum wavefunction as a probability amplitude is not unanimously accepted by the scientific community, the derived meaning of the WDF as a quasi-probability (or probability amplitude) of finding a quantum particle in phase space is not automatically established. This is the role of postulate 3. This probabilistic interpretation is valid for either individual quantum particles or ensembles of quantum particles.

Note that the negative values of the WDF are consistent with its interpretation as the probability amplitude of finding a quantum particle in phase space since not the WDF itself but averages of it over phase space regions indicate the probability; these are always positive as long as the average is performed over regions of phase space not smaller than a quantum blob.

**Postulate 4**: It is not possible to localize a quantum particle in a phase space region smaller than a quantum blob.

This postulate is implicitly assumed in standard quantum mechanics. It is however, acknowledged as the basic postulate of the phase space formulation of quantum mechanics presented in this paper. It is equivalent to stating that the WDF cannot be localized in a phase space region smaller than a quantum blob. However, sub-Planck structures of the WDF can exist in phase space, these representing the sensitivity of the quantum state to perturbations [16].

A quantum blob is defined as any admissible subset of the phase space region with a projection area $A$ on any of the conjugate planes $q_j$, $p_j$ equal to $h/2$. This minimum value of the projection area is a direct consequence of Heisenberg's uncertainty relation. For $n = 1$ a quantum blob is a phase space area equal to $h/2$. The quantum blob is canonical invariant and its definition does not depend on the number of particles, i.e. on the dimension of the phase space [14]. Quantum blobs can have arbitrary shapes and sizes.

The definition of the quantum blob is based on the principle of the "symplectic camel" [14], which states that if a region of phase space $V$, for any number of degrees of freedom $n$, contains a ball with an original area $A$, the area of the projection of the phase space volume $V$ on any of the conjugate planes $q_j$, $p_j$ does not decrease beyond $A$ provided that $V$ is deformed by canonical transformations only. This theorem, which was first proven by Gromov (see [14, 17-18] and the references therein), is a consequence of symplectic topology, and implies that the phase space cross-section defined by conjugate coordinates such as $q_j$ and $p_j$ cannot shrink to zero during a Hamiltonian process, i.e. there is a minimum cross-sectional area within a given volume. The symplectic camel theorem, known also as the non-squeezing theorem, originates from the fact that (for $n > 1$) Hamiltonian flows characterized by symplectic transformations are not simple volume-preserving transformations in phase space, for which the Jacobian matrix has unit determinant, but satisfy more stringent conditions (see [18], 2002). The symplectic camel theorem has been used to develop a rigorous theory of semiclassical mechanics in phase space [18] and, in particular, to correctly predict the semi-classical ground energy levels [17].

Postulate 4 emphasizes the true significance of Planck's constant as determining the "phase space quantum", i.e. the minimum value of the phase space area (or of the physical action) necessary for the existence of a quantum state. It plays in this formulation of quantum mechanics the same role that another universal physical constant – the light speed $c$ – plays in relativity. It is worthwhile mentioning that the original introduction of Planck's constant as proportionality factor between the energy of a light quantum (photon) and its frequency doesn't say a great deal about its significance. On the other hand, the Heisenberg's uncertainty relations between any pair of conjugate variables, for example between the $j$th components of the coordinate and momentum vectors

$$\Delta q_j \Delta p_j \geq \hbar / 2,$$ (7)

provide a better illustration of the meaning of Planck's constant, since $\hbar = h / 2\pi$ does no longer appear on the right hand side of (7) as just a proportionality factor between different physical variables.

Postulate 4 also emphasizes the conceptual significance of the phase space formulation of quantum mechanics. The variances of the conjugated coordinate and momentum operators in (7) act as multiplication with the corresponding variable and scaled partial derivative in both position and momentum representations of the quantum state. The actions of these operators become independent only in the phase space formulation of quantum mechanics, which fully reveals the significance of Planck's constant as determining the minimum phase space area in which a quantum state can be localized.

Two important corollaries follow from postulate 4:

**Corollary 1**: The phase space extent of a quantum particle is smaller than a quantum blob.

This corollary follows immediately from postulates 3 and 4.

**Corollary 2**: A quantum state cannot be prepared in an eigenstate of $\boldsymbol{q}$ or $\boldsymbol{p}$.

The possibility of preparing a quantum state in an eigenstate of $\boldsymbol{q}$ and $\boldsymbol{p}$ would imply that the quantum state (the WDF) is a $\delta$-function localized on a line parallel to the $\boldsymbol{q}$ or $\boldsymbol{p}$ axis in phase space, situation forbidden by the symplectic camel principle since a line has no area and therefore it cannot represent the state of a quantum system. This consequence is stronger than the common assumption that the impossibility to localize a quantum state in a phase space region smaller than a quantum blob means only that the position and the momentum cannot be simultaneously measured. Corollary 2, which is a straightforward conclusion of a reasoning based on phase space concepts and the uncertainty relation has not been, according to the author's knowledge, properly acknowledged by the scientific community, although indications of this fact are known: the eigenstates of the position or momentum operators are not even members of the Hilbert space. (This fact doesn't seem to be properly emphasized in quantum mechanical textbooks; see [12] as reference.) This apparent mathematical anomaly can only be understood in a phase space formulation; it should not be inferred from it that position or momentum operators are not useful. Only their eigenstates, which are $\delta$-functions, cannot represent quantum states; on the other hand, the position and momentum representations of quantum states are well defined in the (classical) phase space.

**Postulate 5**: In the WDF phase space the expectation value of an arbitrary operator $\hat{A}(\hat{q}, \hat{p})$ can be calculated as in the classical phase space, i.e. as

$$\langle \hat{A}(\hat{q}, \hat{p}) \rangle = \int A(q, p)W(q, p)dqdp \,, \tag{8}$$

where $A(q, p)$ is the scalar function obtained by replacing the position and momentum operators in $\hat{A}(\hat{q}, \hat{p})$ with scalar variables $q$ and $p$.

This postulate allows us to demonstrate that the phase space area occupied by the WDF of a

quantum wavefunction that characterizes a physically realizable state attains its minimum

value for a Gaussian wavefunction; a normalized one-dimensional Gaussian wavefunction

$\Psi(q) = (\pi q_0^2)^{-1/4} \exp(-q^2 / 2q_0^2)$ with a spatial extent $q_0$ has a WDF

$W(q, p) = (2/h)\exp(-q^2/q_0^2 - p^2 q_0^2/\hbar^2)$, the variances of the position and momentum

operators being, according to (8), $\Delta q = (\langle \hat{q}^2 \rangle - \langle \hat{q} \rangle^2)^{1/2} = q_0 / 2^{1/2}$ and

$\Delta p = (\langle \hat{p}^2 \rangle - \langle \hat{p} \rangle^2)^{1/2} = \hbar /(2^{1/2} q_0)$, respectively. In this case $\Delta q \Delta p = \hbar / 2$ and the

Heisenberg's uncertainty relation is satisfied at its minimum value.

**Postulate 6**: The WDF evolution law follows directly from the Schrödinger equation satisfied by the wavefunction or from the von Neumann equation obeyed by the density operator.

For $n = 1$ and for a classical Hamiltonian $H(q, p) = p^2 / 2m + V(q)$ this evolution law is

$$\frac{\partial W}{\partial t} + \frac{p}{m}\frac{\partial W}{\partial q} - \frac{\partial V}{\partial q}\frac{\partial W}{\partial p} = \sum_{n=1}^{\infty} \frac{(\hbar/2i)^{2n}}{(2n+1)!}\frac{\partial^{2n+1}V}{\partial q^{2n+1}}\frac{\partial^{2n+1}W}{\partial p^{2n+1}} \,. \tag{9}$$

Postulates 5 and 6 are well known from the phase space formulation of quantum mechanics. They have been introduced here for completeness.

**Postulate 7**: Interference between two quantum states occurs if the respective non-overlapping WDFs have common projections along the $q$ or $p$ lines, while transitions between quantum states occur only when their respective WDFs overlap.

These results have been demonstrated in a number of previous publications (see [19-20]) and have been even applied to predict an Aharonov-Bohm-like effect in the momentum space [21]. They follow, in the mathematical sense, directly from the properties of the WDF. I felt the need, however, to elevate these results to the rank of a postulate for their cognitive value. The distinction between interference and transition, most clearly emphasized in phase space, is intuitively illustrated in Fig.1 for $n = 1$.

It is important to note that the PS description of interference [19] explains also the results of delayed-choice experiments [22]. A quantum particle manifests itself as a wave, if its quantum wavefunction interferes with another, or as a particle in the opposite case depending of whether after an incident quantum wavefunction passes through two slits the measuring device (photographic plate or detector) is introduced at a distance from the two slits for which the WDF of the two wavefunctions have or not common projections along $q$. The choice of introducing one measurement device or another after the particle passes through the slits does not influence the nature of the quantum particle but only establishes the distance from the slits at which the detection is performed (and hence establishes the fact that the detection is made when the WDF of the two wavefunctions have or not common projections along $q$).

On the other hand, the understanding of quantum transitions as a phase space overlap between WDF functions of the initial and final state, both representing phase space probabilities of quantum particles renders the uncomfortable concept of quantum jump unnecessary. This phase space overlap, which determines a transition with the probability

$$P_{12} = h^n \int W_1(\boldsymbol{q}, \boldsymbol{p}) W_2(\boldsymbol{q}, \boldsymbol{p}) d\boldsymbol{q} d\boldsymbol{p} \qquad (10)$$

where $W_1$ and $W_2$ are, respectively, the WDFs of the initial and final states can be viewed as a result of phase space mismatch between different regions. The quantum particle itself does not make a jump; the WDF changes due to interaction, for example, and the incident quantum particle may or may not follow this change depending if its phase space area of possible localization fits or not the new area imposed by interaction. An analytical expression of $P_{12}$ for time-dependent interactions can be found in [20]. The expression of this transition probability and its meaning as phase space overlap should be contrasted to the phase space treatment of filtering, which is developed in the following.

> **Postulate 8**: The interaction between a quantum system and a measuring device can be treated in phase space irrespective of the nature (quantum or classical) of the measuring device. The result of the interaction depends on the type of the measuring device (filter or detector).

Postulate 8 will be first detailed for the case of a quantum mechanical measuring device that can be described by a WDF. However, it will become apparent that classical measuring apparatus can also be accommodated by this theory as long as they can be characterized by a WDF. Note that the variables of the quantum and classical phase spaces are the same, and therefore a classical measuring apparatus can be formally treated in the same way as a quantum one. The only difference between the two cases is how to calculate the WDF. For sizeable devices it might be difficult to calculate a quantum wavefunction or a WDF from quantum considerations and thus the liberty to use a classical description in which the WDF can be determined much more easily is invaluable. Phase space descriptions of both classical light beams [23] and classical ensembles of particles [24] are well developed so that there is no impediment to treat in phase space a macroscopic measuring device. It is also important to specify that the WDF of a classical state changes its ontologic status: if there are a large number of quantum particles that share the same state, the WDF is no longer a probability distribution but a distribution of the number of component quantum particles in phase space, i.e. it represents the object. WDFs with negative values have long been known and even measured in classical optics, for example (see the review in [23]). According to the interpretation in [15] the negative values of the WDF can even disappear for classical objects if there is no correlation between adjacent quantum blobs.

**Corollary 3**: No measuring apparatus is able to measure an eigenstate of the position or momentum operators.

This corollary follows, for quantum measuring apparatus, from corollary 2 and postulate 8; a quantum mechanical apparatus cannot measure an eigenstate of the position or momentum operators because the corresponding state of the apparatus cannot exist. On the other hand, a macroscopic measuring apparatus is formed from a large number of quantum constituents; if a quantum system cannot occupy a phase space region smaller than a quantum blob, it is certain that a macroscopic measuring apparatus cannot either. More precisely, the projection area $A$ occupied by the classical device on any of the conjugate planes $q_j$, $p_j$ scales as $N^{1/n}h$, where $N$ is the number of quantum states with energy smaller than or equal to the energy shell [14]; $N$ is huge for any sizeable device. Following the same line of reasoning as that in the quantum case it is apparent that the WDF of an extended device cannot be a $\delta$-line and therefore it cannot measure an eigenstate of the position or momentum operators. Similar conclusions hold for operators that depend linearly on position or momentum.

The apparent lack of quantum mechanical effects in the classical realm and hence for a macroscopic measuring apparatus with a large number of degrees of freedom $n$ does not automatically imply that quantum mechanics is not applicable but only that quantum mechanical effects are not noticeable since the phase space uncertainty $h/2$ is much smaller than the projection area $A$ occupied by the classical device on any of the conjugate planes $q_j$, $p_j$.

It is important to stress that measurement device is understood in this context as any device that influences the state of a quantum system; it can be either a filter or a detector, these two cases being treated separately in the following sections. In the first case the incident quantum state characterized by a WDF $W_{in}$ can be filtered in either the coordinate or momentum space by the measuring apparatus that is described by a WDF $W_m$, the outgoing quantum state being characterized in phase space by a WDF $W_{out}$. In the second case the result of the measurement is the squared modulus of the quantum wavefunction that results after the incident wavefunction is filtered in some way by the measuring device.

### 3. Phase space effect of a filtering device

A filtering device is any device that influences the evolution of a quantum state such that the outgoing wavefunction has some "memory" of its original form. A filter influences the results of a subsequent measurement since it alters the incident wavefunction and therefore actively manipulates the result of the measurement.

Let us consider first a quantum filter. If the incident quantum wavefunction $\Psi_{in}(\boldsymbol{q})$ is filtered in the coordinate space by the transmission function $\Psi_m(\boldsymbol{q})$ of a filter, the output wavefunction is $\Psi_{out}(\boldsymbol{q}) = \Psi_{in}(\boldsymbol{q})\Psi_m(\boldsymbol{q})$. This filtering action can be described in phase space as

$$W_{out}(\boldsymbol{q},\boldsymbol{p}) = \int W_{in}(\boldsymbol{q},\boldsymbol{p}')W_m(\boldsymbol{q},\boldsymbol{p}-\boldsymbol{p}')d\boldsymbol{p}', \tag{11}$$

i.e. as a mere multiplication along the coordinate axis in phase space and a convolution along the momentum direction. Similarly, a filter with a transmission function in the momentum space $\overline{\Psi}_m(\boldsymbol{p})$ transforms an incident quantum wavefunction in the momentum representation $\overline{\Psi}_{in}(\boldsymbol{p})$ into $\overline{\Psi}_{out}(\boldsymbol{p}) = \overline{\Psi}_{in}(\boldsymbol{p})\overline{\Psi}_m(\boldsymbol{p})$, transformation that can be represented in phase space as

$$W_{out}(\boldsymbol{q}, \boldsymbol{p}) = \int W_{in}(\boldsymbol{q}', \boldsymbol{p}) W_m(\boldsymbol{q} - \boldsymbol{q}', \boldsymbol{p}) d\boldsymbol{q}'. \tag{12}$$

The convolution is now performed along the coordinate axis of the phase space, the transformation along the momentum axis being a simple multiplication.

A more general filtering process can generate an output wavefunction of the form

$$\Psi_{out}(\boldsymbol{q}) = h^{-n/2} \int \Psi_{in}(\boldsymbol{q}')\Psi_m(\boldsymbol{q} - \boldsymbol{q}') \exp(i\boldsymbol{p}_0\boldsymbol{q}'/\hbar)d\boldsymbol{q}'; \tag{13}$$

This filtering process can be represented in phase space as

$$W_{out}(\boldsymbol{q}, \boldsymbol{p}) = \int W_{in}(\boldsymbol{q}', \boldsymbol{p} - \boldsymbol{p}_0) W_m(\boldsymbol{q} - \boldsymbol{q}', \boldsymbol{p}) d\boldsymbol{q}', \tag{14}$$

the similar expression

$$W_{out}(\boldsymbol{q}, \boldsymbol{p}) = \int W_{in}(\boldsymbol{q} - \boldsymbol{q}_0, \boldsymbol{p}') W_m(\boldsymbol{q}, \boldsymbol{p} - \boldsymbol{p}') d\boldsymbol{p}' \tag{15}$$

describing the filtering process

$$\overline{\Psi}_{out}(\boldsymbol{p}) = h^{-n/2} \int \overline{\Psi}_{in}(\boldsymbol{p}')\overline{\Psi}_m(\boldsymbol{p} - \boldsymbol{p}') \exp(-i\boldsymbol{q}_0\boldsymbol{p}'/\hbar)d\boldsymbol{p}'. \tag{16}$$

Note that in all phase space expressions (11), (12), (14) and (15) only the WDF of the filter appears; it is irrelevant if this is calculated or not with the use of quantum mechanical principles since the WDF of quantum and classical physics are the same. Therefore, the effect of a classical filter on a quantum state can equally be described by equations (11), (12), (14) and (15).

A filtering process is different from interference since the WDFs of the incident quantum state and the filtering device must (at least partially) overlap. It differs also from transition since the filtering is not just a phase space overlap; moreover, in quantum transitions there is no intermediate filter. There is only the incident wavefunction and an outgoing wavefunction, totally different from the first. Unlike filtering, quantum transition has no "memory" of the original wavefunction except for transition probability.

### 3.1. Examples

It is worthwhile to exemplify the influence of a filter in the spatial domain for the case $n = 1$. Let us consider that the input wavefunction $\Psi_{in}(q) = (\pi q_i^2)^{-1/4} \exp(-q^2 / 2q_i^2)$ is a Gaussian with a spatial extent $q_i$ and that the quantum filter is a slit with a normalized transmission function $\Psi_m(q) = (\pi q_m^2)^{-1/4} \exp(-q^2 / 2q_m^2)$. The corresponding WDFs, calculated as exemplified in relation with postulate 5, are represented in Fig.2 for the case $q_i > q_m$; darker areas correspond to higher values of the WDFs. Note that, since the WDF of a Gaussian is localized in a phase space area equal to $h / 2$, the extent of the WDF on the $q$ axis is inversely proportional to the extent on the $p$ axis. The WDF of the output state is given, according to (11), by

$$W_{out}(q, p) = \frac{2}{h\sqrt{\pi(q_i^2 + q_m^2)}} \exp\left(-\frac{q^2}{q_i^2} - \frac{q^2}{q_m^2}\right) \exp\left(-\frac{p^2}{\hbar^2} \frac{q_i^2 q_m^2}{q_i^2 + q_m^2}\right) \tag{16}$$

From (16) it follows that in the limit $q_i >> q_m$ the WDF of the output state is a scaled version of the WDF of the filter, while if $q_i << q_m$ the filter does not significantly influence the input state. These results are intuitive when we view the filtering process in the coordinate representation, but are somehow counterintuitive when looking at Fig.2. A simple filtering in phase space along both $q$ and $p$ would result in an output WDF given by the intersection of the WDFs of the input state and filter; however, the convolution transform along $p$ results in an output WDF that can have an extent along $p$ much larger than the extent of the intersection of the two WDFs. A similar phase space approach towards the study of the quantum coordinate measurement with a hard slit was taken in [13] although the form (11) does not explicitly appear in this reference.

The result that if $q_i << q_m$ the filter does not significantly influence the input state becomes extremely suggestive if we interpolate it in the classical limit of both states and apparatus. As we have already mentioned a classical object can also be represented in phase space by a probability distribution or even by a non-positive-valued WDF if it is a classical field. A filter cannot let a classical object pass unless it has larger dimensions. According to (16), a slit (much) larger than the object leaves it practically unchanged.

Another interesting example is that of a slit put in front of a cat-like quantum state, which is characterized by the one-dimensional wavefunction

$$\Psi_{in}(q) = N\left[\exp\left(-\frac{(q-d)^2}{2q_i^2}\right) + \exp\left(-\frac{(q+d)^2}{2q_i^2}\right)\right],\tag{17}$$

where the normalization constant $N = (4\pi q_i^2)^{-1/4}[1+\exp(-d^2/q_i^2)]^{-1/2}$. The WDF of a cat state, given by

$$W_{in}(q,p) = \frac{\exp(-p^2q_i^2/\hbar^2)}{h[1+\exp(-d^2/q_i^2)]}\left[\exp\left(-\frac{(q-d)^2}{q_i^2}\right) + \exp\left(-\frac{(q+d)^2}{q_i^2}\right) + 2\exp\left(-\frac{q^2}{q_i^2}\right)\cos\left(\frac{2dp}{\hbar}\right)\right],$$

$$\tag{18}$$

consists (for a sufficiently large separation $d$) of two outer terms, which represent the WDFs of the individual Gaussian constituents, and of an interference term (the last term in (18)). Fig.3 shows the WDF of a cat state for $d = 4q_i$; darker areas represents higher values of the WDF. Note that the middle, oscillatory interference term has larger amplitude than the outer terms and attains its maximum value at the phase space origin $q = 0$, $p = 0$.

If this WDF is filtered by a quantum Gaussian slit that is centered at $q = D$ the output WDF calculated according to (11) is given by

$$
\begin{aligned}
W_{out}(q, p; D) = K \exp\left(-\frac{(q-D)^2}{q_m^2}\right) \exp\left(-\frac{p^2}{\hbar^2}\frac{q_i^2 q_m^2}{q_i^2 + q_m^2}\right) \\
\times \left[ \exp\left(-\frac{(q-d)^2}{q_i^2}\right) + \exp\left(-\frac{(q+d)^2}{q_i^2}\right) + 2\exp\left(-\frac{q^2}{q_i^2}\right)\exp\left(-\frac{d^2}{q_i^2 + q_m^2}\right)\cos\left(\frac{2dp}{\hbar}\frac{q_m^2}{q_i^2 + q_m^2}\right)\right]
\end{aligned}
$$

(19)

This formula tells us that, as above, the filter does not significantly influence the input state if $q_i << q_m$ and $d << q_m$. However, the different terms in $W_{in}$ are differently transformed as a result of the filtering if $q_i \cong q_m$. In our specific example, for $d = 4q_i$, a slit with $q_m \cong q_i$ cannot filter all three terms of the incident WDF. To study the influence of the filtering on these terms it is instructive to modify the off-axis filter position $D$ across the width of the cat-like state. The resulting function $W_{out}(q, 0; D)$, displayed in Fig.4, shows that a slit with $q_m = q_i$ really filters the outer terms only; only the contributions of the incident WDF that correspond to the probability of finding particles and not the interference term are "sensed" by the slit. The interference term, which does not correspond to any significant probability of finding quantum particles but only signals the potentiality of interference, is ignored by the slit due to the convolution along $p$ performed in the filtering process although its amplitude in Fig.3 is larger than that of the outer terms. This example illustrates better than any other that phase space filtering refers to filtering of the regions in which the probability of finding the incident quantum wavefunction (in coordinate or position representation) is significant. This is to be expected from our definition $\Psi_{out}(q) = \Psi_{in}(q)\Psi_m(q)$ of the slit action. Phase space representation of quantum states must be handled with care in order to distinguish between "true" terms and "phony" interference terms; these have different significances and are essential in different cases. A similar conclusion was established in connection with phase space interference [19].

### 3.2. Observation

Can a filter act in a more sophisticated manner than just a transmission function in coordinate or momentum representations? Perhaps it can, but it is quite difficult to imagine the action of such a filter. Coordinate and momentum are complementary variables; the external motion of quantum states can be described only in terms of these. Other internal quantum variables manifest their existence also in the coordinate or momentum space; it is this possibility to discern various internal quantum states at different position or momentum coordinates after passing through appropriately designed set-ups that make internal quantum variables observable. For example, different spin component values are spatially separated due to the different deflection directions from a Stern-Gerlach experimental set-up. An introduction of internal quantum variables in the definition of the quantum WDF has not been carried out up to now but it may be considered necessary as the phase space formulation of quantum mechanics becomes more widespread. Up to that moment the filtering theory in phase space presented in this paper covers the majority of practical situations.

### 4. Phase space effect of a detector

A convolution of the WDF of the filtered quantum state along both coordinate and momentum directions cannot generally represent a WDF of a quantum wavefunction since

$$\int W_{in}(\boldsymbol{q'}, \boldsymbol{p'}) W_m(\boldsymbol{q} - \boldsymbol{q'}, \boldsymbol{p} - \boldsymbol{p'}) d\boldsymbol{q'} d\boldsymbol{p'} = h^{-n} \mid \int \Psi_{in}(\boldsymbol{q'}) \Psi_m^*(\boldsymbol{q} - \boldsymbol{q'}) \exp(-i\boldsymbol{p}\boldsymbol{q'}/\hbar) d\boldsymbol{q'} \mid^2 \qquad (20)$$

is a positive definite function. However, it can represent the result of a measurement of the input quantum state with a quantum detection device that filters it in both position and momentum coordinates. This phase space representation of the quantum detection process is not new [25, 26]. However, as advocated in the previous section, the left hand side of (20) describes equally well the result of detection of a quantum state with a classical detector characterized by a WDF $W_m$.

Note that, according to (10), the detection process expressed by (20) is a generalized transition process. The quantum particle performs a transition from the input state to the detector, its presence being observed by a "click" or some other manifestation. Even at detection the measured phase space distribution of quantum particles is not independent of the inherent filtering process performed by any detector. It is not possible to detect quantum states without perturbing them.

## 5. Discussions and conclusions

We have developed a quantum mechanical phase space formalism that incorporates not only an established mathematical formalism but also ontological interpretations of the physical reality, focusing on the characterization in phase space of the result of filtering and detection upon an incident quantum state. Although some of the mathematical formulations presented in this paper are well known, the phase space treatment of the filtering process as well as the axiomatic form of the theory and the ontological interpretation are novel. At the present level of the development of the phase space formalism of quantum mechanics there appear to be no differences between the predictions of the theory in this paper and standard quantum mechanics as long as quantum concepts are used. However, differences exist, one of the most important being that of the possibility of introducing an equal treatment of quantum and classical measuring apparatus. A subsequent development of the theory, which is only drafted here, might reveal other differences.

Although only the measuring apparatus can be classical in the form of the theory developed here, all the results, including the description of the input state, can be extended to the classical domain. The particular form of the theory was expressed with the goal of trying to bring an insight in the unsolved quantum mechanical measurement problem; the classical measurement problem is far less (if any) controversial. The quantum-classical correspondence is most relevant in the phase space formulation, since both theories have similar mathematical formulations (see [27] for a detailed treatment of quantum-classical correspondence in phase space). The fact that the conversion from quantum to classical terminology does not affect the description of a measuring apparatus if the dynamical variable is an operator of the Weyl-Wigner type has already been shown in reference [28].

A few remarks on the relation between the formalism presented in this paper and other interpretations of quantum mechanics, especially of the measurement process, are in order. In the phase space formulation of the quantum theory of measurement it is no need to assume any reduction of the wavefunction of the initial quantum state, since this wavefunction does not need to be decomposed in eigenstates of some operator. The problem changes from finding the eigenstates to that of how to design the filter/detector (what WDF should it have?) to gather information about a specific property. Therefore, there is no need for a cut in the measurement process between the quantum system and the (classical) measuring device; this is welcome since the location of the cut in the Copenhagen interpretation is to a large extent arbitrary. As discussed above, even classical devices can be accommodated by the present phase space theory if their WDF can be deduced. The subsequent theory of interference, measuring and transition probabilities in phase space can be applied irrespective of the manner in which the WDF are determined.

The theory presented in this paper responds to most of desiderates of the so-called Ithaca interpretation of quantum mechanics [29]. Namely, it is based on the fact that (i) quantum mechanics describes an objective reality, (ii) it is based on the notion of objective probability, (iii) describes individual systems, not just ensembles, (iv) does not have to invoke interactions with environment (or the existence of a classical domain) when describing small isolated systems, and, most important (v) the concept of measurement plays no fundamental role. Although not explicitly stated the sixth desiderate, that (vi) objectively real internal properties of an isolated individual system do not change when another non-interacting system is perturbed, follows implicitly from the theory presented in this paper.

The probabilistic interpretation of the measurement results, in which the WDF is seen as the probability amplitude of a quantum particle (or particles) that cannot be confined to a phase space region smaller than a quantum blob, can explain the so-called wave-particle duality. As discussed with regard to the phase space interpretation of delayed choice experiments the "choice" of wave or particle is done by the place where the detection device is made and by the type of the detection device; since the detection process is local the outcome of the result is only determined by the form of the WDF of the total system (two quantum states) at that particular place (and moment) of detection. All possible types of interactions: interference, filtering, transitions and detection have different "signatures" in the phase space treatment and therefore this approach to quantum mechanics is best suited.

**Figure captions**

Fig.1 (a) Two states with individual WDFs $W_1$ and $W_2$ that have common projections along the $q$ and $p$ axes interfere along both spatial and angular coordinates. (b) Transitions are expected to occur if the individual WDFs overlap at least partially.

Fig.2 WDFs of an incident Gaussian wavefunction, $W_{in}$, and a filter with a Gaussian transmission, $W_m$, aligned with respect to the incident wavefunction.

Fig.3 WDF of a cat-like quantum state. Darker areas represent larger values of the WDF.

Fig.4 Filtering effect of a narrow misaligned, off-axis Gaussian slit upon a cat-like state. Darker areas represent larger values of the WDF

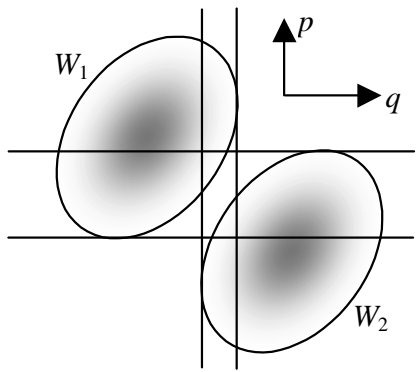

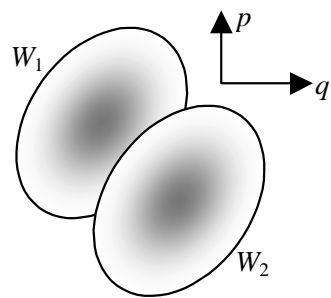

Figure 1

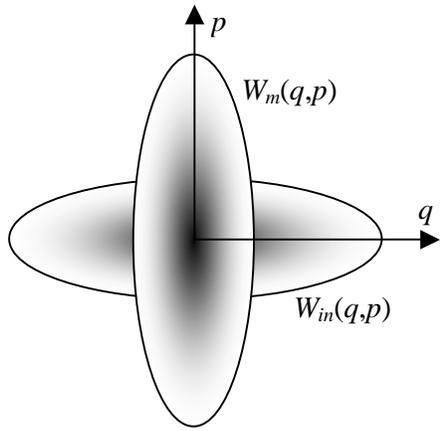

Figure 2

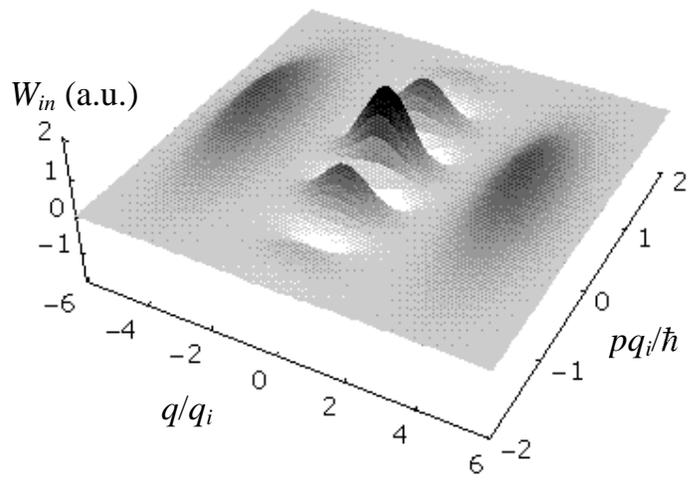

Figure 3

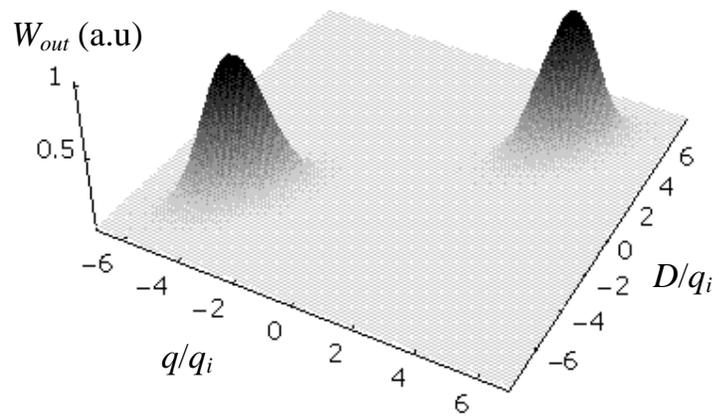

Figure 4